\documentclass[aps,prl,twocolumn,preprintnumbers,amsmath,amssymb,superscriptaddress]{revtex4-1}

\usepackage{graphicx}% Include figure files
\usepackage{bm}% bold math
\usepackage{color}
\usepackage{hyperref}

\usepackage{t1enc}
\usepackage[polish,english]{babel}
\usepackage[cp1250]{inputenc}
\usepackage{polski}
\usepackage[T1]{fontenc}
\begin{document}

\title{Interplay of the Kondo Effect and
Spin-Polarized Transport in Magnetic Molecules, Adatoms and Quantum Dots}

\author{Maciej Misiorny}
 \email{misiorny@amu.edu.pl}
\affiliation{Faculty of Physics, Adam Mickiewicz University,
61-614 Pozna\'{n}, Poland}

\author{Ireneusz Weymann}
\affiliation{Faculty of Physics, Adam Mickiewicz University,
61-614 Pozna\'{n}, Poland} \affiliation{Physics Department, ASC,
and CeNS, Ludwig-Maximilians-Universit\"at, 80333 Munich, Germany}

\author{J\'{o}zef Barna\'{s}}
\affiliation{Faculty of Physics, Adam Mickiewicz University,
61-614 Pozna\'{n}, Poland} \affiliation{Institute of Molecular
Physics, Polish Academy of Sciences, 60-179 Pozna\'{n}, Poland}

\date{\today}

\begin{abstract}
We study the interplay of the Kondo effect and spin-polarized
tunneling in a class of systems exhibiting uniaxial magnetic
anisotropy.
%, such as magnetic molecules, magnetic adatoms, or
%quantum dots coupled to a single localized magnetic moment.
Using
the numerical renormalization group method we calculate the
spectral functions and linear conductance in the Kondo regime. We
show that the exchange coupling between conducting electrons and
localized magnetic core generally leads to suppression of the
Kondo effect. We also predict a nontrivial dependence of the
tunnel magnetoresistance on the strength of exchange coupling and
on the anisotropy constant.
\end{abstract}


%72.25.-b Spin polarized transport
%73.23.-b Electronic transport in mesoscopic systems
%75.50.Xx  Magnetic devices: molecular magnets
%75.60.Jk Magnetization reversal mechanisms
%85.75.-d Magnetoelectronics; spintronics: devices exploiting spin
%polarized transport or integrated magnetic fields

\maketitle

%=========================================================================================================================================

Incorporating single atoms or molecules into nanoelectronic
devices is a very promising challenge, particularly for
information storage and processing
technologies~\cite{Joachim_Nature408/00, *Tans_Nature393/98,
*Park_Nature407/00, *Piva_Nature435/05, *Heath_PhysToday56/03,
*Bogani_NatureMater7/08,Green_Nature445/07,Mannini_NatMater8/09,Loth_NatPhysics6/09}.
Owing to recent achievements in experimental techniques, measuring
transport through individual natural and artificial (quantum dots)
atoms and molecules has become feasible. Among various nanoscopic
systems, magnetic atoms of spin $S>1/2$ (like Fe, Co or
Mn)~\cite{Hirjibehedin_Science317/07,Otte_NatPhysics4/08,Loth_NatPhysics6/09}
and single-molecule magnets (SMMs), both exhibiting magnetic
anisotropy ~\cite{Gatteschi_book,Heersche_PRL96/06,
*Jo_NanoLett6/06, *Voss_PRB78/08, *Zyazin_NanoLett10/10}, have attracted much interest
from both the fundamental as well as application points of view.
In particular, it has been suggested that magnetic states of SMMs can be
controlled by spin-polarized current~\cite{Elste_PRB73/06,
*Timm_PRB73/06, *Misiorny_PRB75/07, *Misiorny_PSS246/09,
*Delgado_PRL104/10,Misiorny_PRB79/09,Lu_PRB79/09}, i.e. a
sufficient current pulse can switch the magnetic moment between
two low energy states (the model applies also to electric control
of magnetic atoms). This has been recently proven experimentally
in the case of magnetic adatoms~\cite{Loth_NatPhysics6/09} using
spin-polarized STM technique. Furthermore, a SMM weakly coupled to
two nonmagnetic metallic electrodes was shown to act as a spin
filter~\cite{Barrazalopez_PRL102/09}, while coupled to electrodes
with different spin polarizations reveals features typical of a
spin diode~\cite{Misiorny_EPL89/10}. In the strong coupling limit,
on the other hand, the Kondo correlations play a significant role
and anomalous features of transport become revealed at low
temperatures. As shown experimentally, the magnetic anisotropy in
such systems can be used to tune the Kondo
effect~\cite{Otte_NatPhysics4/08,Parks_Science328/10}. Using
break-junction technique, Parks {\it et
al.}~\cite{Parks_Science328/10} were able to tune the anisotropy
constant and therefore modify the energy spectrum responsible for
the Kondo state, which in turn resulted in a crossover from the
fully screened to underscreened Kondo effect. The Kondo phenomenon
in transport through SMMs has been a subject of current interest,
but the research so far was mainly focused on the interplay of the
Kondo effect and quantum tunneling of SMM's
spin~\cite{Romeike_PRL96I/06,
*Romeike_PRL97/06, *Leuenberger_PRL97/06, *Gonzalez_PRL98/07, *Gonzalez_PRB78/08}.

To our knowledge, the physics of Kondo correlations in
spin-polarized transport through SMMs and/or magnetic adatoms is
still rather unexplored. This problem is therefore addressed in
the present Letter. We consider a situation when transport occurs
{\it via} a local orbital of the system (orbital of a SMM, adatom,
or a quantum dot), which is coupled to electrodes and additionally
exchange-coupled to the corresponding magnetic core. The role of
magnetic anisotropy in conductance and tunnel magnetoresistance
(TMR) is also analyzed. We show, that the exchange coupling to the
corresponding magnetic core generally suppresses the Kondo effect.
Similarly, the magnetic anisotropy has a significant impact on the
conductance and TMR in the Kondo regime.

\begin{figure}[t]
  \includegraphics[width=0.9\columnwidth]{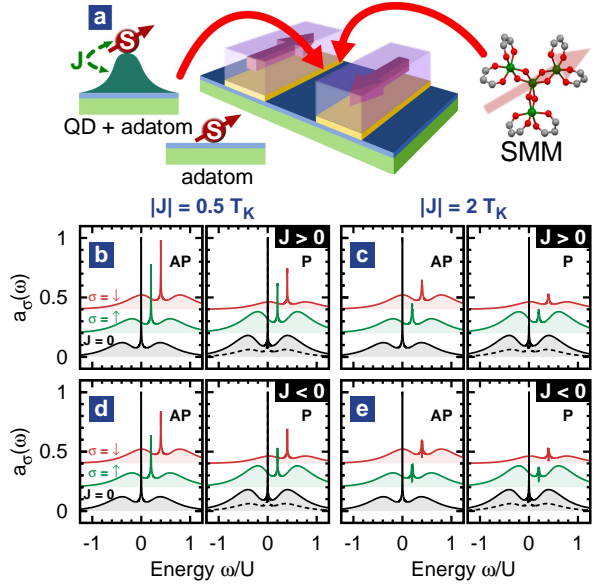}
  \caption{\label{Fig:1} (Color online)
  (a) Schematic of a molecular quantum dot (magnetic molecule, magnetic adatom, or quantum dot
  exchange-coupled to a magnetic moment) coupled to two ferromagnetic electrodes.
  (b)-(e) Normalized spin-resolved orbital level (OL)
  spectral functions $a_\sigma(\omega)=\pi\Gamma_\sigma A_\sigma(\omega)$
  in the \emph{antiparallel} (left) and \emph{parallel} (right part of each figure)
  magnetic configuration of electrodes for two different
  values of the exchange  parameter $J$.
  The bottom curves (black) in each plot correspond to $J=0$,
  with the solid line referring to $\sigma=\uparrow$ and
  the dashed one to $\sigma=\downarrow$. To increase readability, for $J\neq0$
  the curves for $\sigma=\uparrow$ ($\sigma=\downarrow$) are shifted up and right by 0.2 (0.4).
  The parameters are: $S=2$, $U=1$ meV, $\varepsilon/U=-0.5$,
  $\Gamma/U\approx0.075$, $D/U\approx1.7\cdot10^{-4}$, and $p=0.5$.}
\end{figure}

\emph{Model --} We consider a generic model that includes
essential features of such quantum objects like SMMs, magnetic
adatoms, and quantum dots exchange coupled to local magnetic
moments, see Fig.~\ref{Fig:1}(a). These systems are referred to as
magnetic quantum dots (MQDs). We assume that MQD is coupled to
ferromagnetic leads whose magnetizations can form either {\em
parallel} (P) or {\em antiparallel} (AP) configuration, while MQD's
magnetic easy axis is collinear with magnetic moments of the
leads. We also assume that only a single orbital level (OL) of a
MQD (lowest unoccupied molecular orbital of a SMM, atomic or
quantum dot level) is directly coupled to the leads, while the
corresponding magnetic core is coupled to the leads indirectly
\emph{via} exchange coupling to electrons in the local OL. The
corresponding Hamiltonian reads~\cite{Elste_PRB73/06,
*Timm_PRB73/06, *Misiorny_PRB75/07, *Misiorny_PSS246/09,
*Delgado_PRL104/10,Misiorny_PRB79/09}
\begin{equation}\label{Eq:1}
    \mathcal{H}_\textrm{MQD} = -D S_z^2
    + \sideset{}{_\sigma}\sum \varepsilon\,n_\sigma
    + U\, n_\uparrow n_\downarrow
    -J \textbf{s}\cdot\textbf{S},
\end{equation}
with $S_z$ denoting the $z$th component of the MQD's core spin
operator $\bm{S}$, and $D$ being the uniaxial anisotropy constant
of the MQD. The operator $n_\sigma=c^\dag_\sigma c_\sigma^{}$,
where $c_\sigma^\dag (c_\sigma^{})$ creates (annihilates) an
electron of energy $\varepsilon$ in the OL, whereas $U$ denotes
the Coulomb energy of two electrons occupying the OL. Energy (and
therefore occupation) of this level can be controlled by an
external gate voltage. Finally, the last term accounts for
exchange coupling between the MQD's magnetic core and electrons in
the OL, with $\textbf{s}=(1/2)\sum_{\sigma\sigma'}c_\sigma^\dag
\bm{\sigma}_{\sigma\sigma'}^{} c_{\sigma'}^{}$  and
$\bm{\sigma}\equiv(\sigma^x,\sigma^y,\sigma^z)$ denoting the Pauli
matrices.

\emph{Method --} To accurately address the problem of transport
through MQDs in the strong coupling regime, we use the
\emph{Wilson's numerical renormalization group} (NRG) method
\cite{Wilson_RMP47/75}. The NRG Hamiltonian of the full system can
be then written as~\cite{Hewson_book, *Bulla_RevModPhys80/08}
    \begin{align}\label{Eq:2}
    \mathcal{H}=\mathcal{H}_\textrm{MQD} &+ \sideset{}{_\sigma}
    \sum \sqrt{\Gamma_\sigma/(\pi\rho)}
    \big[c_\sigma^\dag f_{0\sigma}^{}+f_{0\sigma}^\dag c_\sigma^{}\big]
    \nonumber\\
    &+\sideset{}{_{\sigma,n=0}^{\infty}}\sum t_n\big[f_{n\sigma}^\dag f_{n+1\sigma}^{}
    +f_{n+1\sigma}^\dag f_{n\sigma}^{}\big],
    \end{align}
where $f_{n\sigma}$ ($f_{n\sigma}^{\dag}$) represents the $n$th
site of the Wilson's chain (last term of $\mathcal{H}$) and $t_n$
denotes the hopping matrix element between neighboring sites of
the chain. The second term of the NRG Hamiltonian stands for the
coupling between the MQD and conduction electrons. Here, we use
the flat band approximation, with the density of states
$\rho=1/(2\mathfrak{D})$, where $\mathfrak{D}$ is the band half-width.
The effect of ferromagnetic leads is then determined by the
hybridization function $\Gamma_\sigma = \Gamma_{\rm L\sigma} + \Gamma_{\rm R\sigma}$~\cite{Martinek_PRL91/03,
*Sindel_PRB76/07}, where $\Gamma_{r\sigma}$
is the coupling to the $r$th lead.
When assuming left-right symmetry, the resultant coupling in the AP magnetic configuration
does not depend on spin,
$\Gamma_{\uparrow(\downarrow)}^\textrm{AP}=\Gamma$,
and the system behaves then as being coupled to nonmagnetic leads.
This is not the case in the
P configuration where,
$\Gamma_{\uparrow(\downarrow)}^\textrm{P}=\Gamma(1\pm p)$, with
$p$ being the spin polarization of the leads and $\Gamma =
(\Gamma_\uparrow+\Gamma_\downarrow)/2$.
By solving Hamiltonian (\ref{Eq:2}) iteratively, we are able to determine
static and dynamic quantities, basically at arbitrary energy~\cite{Bulla_RevModPhys80/08}.
Transport properties of a MQD are then determined from the spectral function of the OL,
$A_\sigma(\omega) = -\frac{1}{\pi}\textrm{Im} \{G^{\rm
R}_\sigma(\omega)\}$, where $G^{\rm R}_\sigma(\omega)$ is the
Fourier transform of the retarded Green's function, $G_\sigma^{\rm
R}(t) =
-i\Theta(t)\langle\{c_\sigma(t),c_\sigma^\dag(0)\}\rangle$.

\begin{figure*}[t]
  \includegraphics[width=2.05\columnwidth]{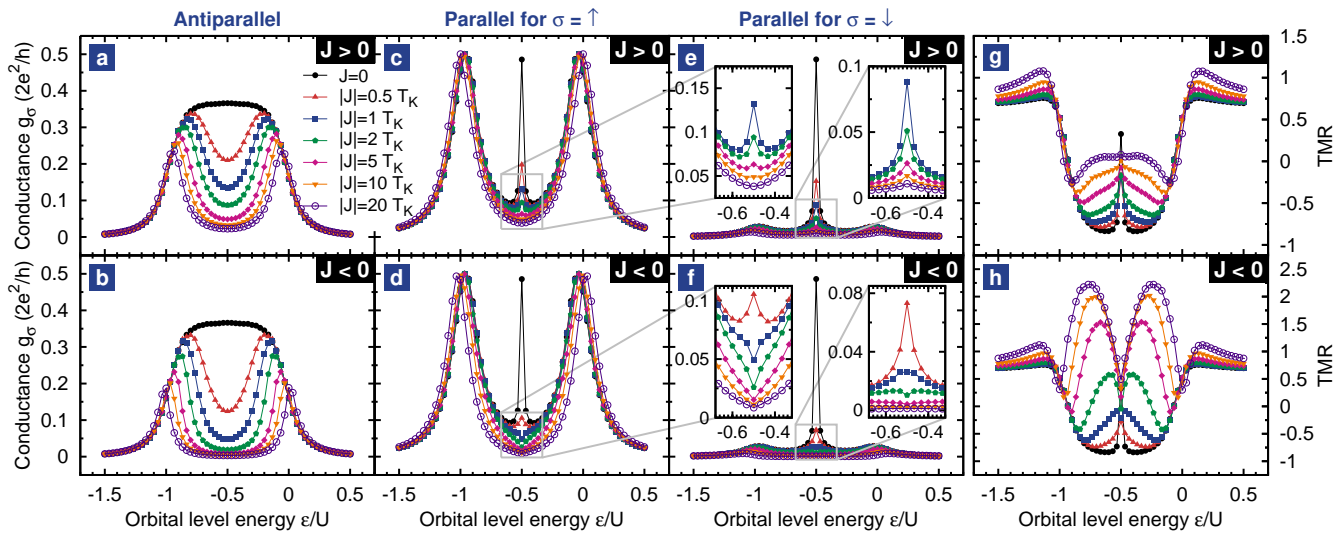}
  \caption{\label{Fig:2} (Color online)
  (a)-(f) Dependence of the linear conductance
  $g_\sigma$ on the OL energy  $\varepsilon$  for
  indicated values of the exchange parameter $J$ in
  the \emph{antiparallel} (a)-(b) and \emph{parallel}
  (c)-(f) magnetic configuration.
  Figures (g) and (h) show the corresponding TMR.
  Top panel represents the case of \emph{ferromagnetic}
  ($J>0$) coupling between electrons in the OL
  and MQD's magnetic core, whereas the bottom one
  corresponds to the \emph{antiferromagnetic} ($J<0$) coupling.
  Parameters as in Fig.~\ref{Fig:1}.}
\end{figure*}

\emph{Spectral functions --} The zero-temperature spin-resolved
spectral functions $A_\sigma(\omega)$ of a singly occupied OL  are
shown in Figs.~\ref{Fig:1}(b)-(e) for the P and
AP configurations and for $\varepsilon=-U/2$. To begin with, we
note that in the AP configuration the effective coupling
is the same for both spin orientations. This is, however, opposite
to the P configuration where  an effective spin
splitting of the OL due to spin-dependent coupling to the leads
(effective exchange field) occurs~\cite{Martinek_PRL91/03}. This exchange
field depends on the position of the OL as $\delta \varepsilon
_{\rm exch} \sim \ln |\varepsilon/(\varepsilon+U)|$. Consequently,
the splitting will generally suppress the Kondo effect, except for
the case of particle-hole symmetry point $\varepsilon=-U/2$ (shown
in Fig.~\ref{Fig:1}) where $\delta \varepsilon _{\rm exch} \to 0$.
Therefore, for $\omega \lesssim T_{\rm K}$, where $T_{\rm K}$ is
the Kondo temperature~\footnote{The system's Kondo temperature
defined as half-width of the total spectral function $\sum_\sigma
A_\sigma(\omega)$ in the AP configuration for
$\varepsilon=-U/2$ is $T_\textrm{K} \approx 0.022 \Gamma $.}, the
spectral functions in both configurations display a clear
Kondo-Abrikosov-Suhl resonance at the Fermi level.

When considering the effect of finite exchange interaction $J$,
one should note that there are two competing interactions
corresponding to the energy scales set by $J$ and
$T_{\textrm{K}}$. When $|J| < T_{\textrm{K}}$, the system then
tends to lower its energy by $T_{\textrm{K}}$ due to formation of
the many-body Kondo state. However, once $|J|
> T_{\textrm{K}}$, it becomes energetically more favorable for an
electron in the OL to hybridize with the core spin $S$ instead of
free electrons in the leads. Thus, the resonant peak at the Fermi
level will not develop in such a case.
In Figs.~\ref{Fig:1}(b)-(e) we show the spectral functions
corresponding to both \emph{ferromagnetic} ($J>0$) and
\emph{antiferromagnetic} ($J<0$) exchange couplings.
It can be clearly seen that the Kondo effect becomes
suppressed by the exchange coupling $J$, and this suppression is
stronger for antiferromagnetic coupling. Actually, full
suppression appears when $|J| \gtrsim T_{\textrm{K}}$.

\begin{figure}[t]
  \includegraphics[width=0.99\columnwidth]{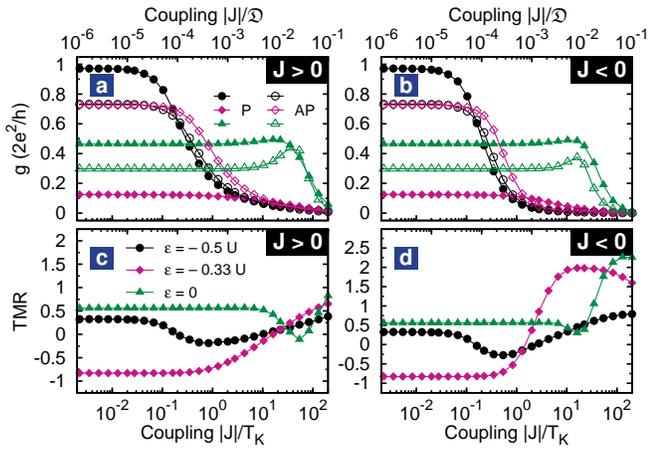}
  \caption{\label{Fig:3} (Color online)
  Conductance $g=\sum_\sigma g_\sigma$ (a,b) and TMR (c,d) in the
  \emph{parallel} (P) and \emph{antiparallel} (AP)
  configurations as a function of the exchange parameter
  $J$ for indicated values of the OL energy $\varepsilon$.
  The dashed vertical line indicates the Kondo temperature.
  Other parameters as in Fig.~\ref{Fig:1}.}
\end{figure}

\emph{Linear conductance --} The spectral functions determine the
linear conductance $g_\sigma$:
$g_{\uparrow(\downarrow)}^\textrm{AP} = (e^2/h)(1-p^2)\pi\Gamma
A_{\uparrow(\downarrow)}^\textrm{AP}(0)$ for the
\emph{antiparallel} configuration, and
$g_{\uparrow(\downarrow)}^\textrm{P}=(e^2/h)(1\pm p)\pi\Gamma
A_{\uparrow(\downarrow)}^\textrm{P}(0)$ for the \emph{parallel}
one, where $A^{\rm P/AP}_\sigma(0)$ is the spectral function in
the P/AP configuration for $\omega=0$. The linear
conductance as a function of the OL energy is shown in
Figs.~\ref{Fig:2}(a)-(f) for different values of $J$. First, for
$|J| \ll T_{\rm K}$ we recover results known for single-level
quantum dots~\cite{Choi_PRL92/04}. For the AP magnetic
configuration one observes then an enhanced conductance due to the
Kondo state when the orbital level is singly occupied. The
conductance is then given by,
$g_{\uparrow(\downarrow)}^\textrm{AP} = (e^2/h)(1-p^2)$, i.e. it
is reduced by a polarization-dependent factor $(1-p^2) = 3/4$, as
compared to the conductance quantum, see Figs.~\ref{Fig:2}(a,b)
and also Figs.~\ref{Fig:3}(a,b) for $|J|/T_{\textrm{K}} \ll 1$. In
the P configuration, in turn, the Kondo effect is
suppressed due to effective spin splitting of the
OL~\cite{Martinek_PRL91/03}, except for $\varepsilon = -U/2$ where
a sharp peak occurs and the Kondo effect is restored.

When $J\neq0$, the Kondo anomaly in conductance becomes gradually
suppressed with increasing $|J|$, and this suppression is faster
for $J<0$ than for $J>0$. This behavior is a consequence of
 a difference in quantum states taking part in
formation of the Kondo state for $J<0$ and $J>0$. First, the
ground state energy $E_{\rm GS}$ of a singly-occupied bare MQD is
lower for $J<0$ than for $J>0$, $E_{\rm GS}^{\rm AF} < E_{\rm
GS}^{\rm F}$, while the energies of virtual states corresponding
to empty and doubly occupied OL  are independent of $J$.
Second, the singly occupied MQD's ground state for $J<0$ involves
a superposition of both electronic spin states `up' and `down', which is not the case for $J>0$.
Consequently, the cotunneling processes driving the Kondo effect for $J<0$ are suppressed more
effectively than for $J>0$, and this behavior is clearly visible
in the linear conductance, see Figs.~\ref{Fig:2}(a,b). When the
Kondo peak becomes suppressed, only two main resonances appear in
the conductance, whose position depends on $J$: for $J>0$ they
appear at $\varepsilon = JS/2$ and $\varepsilon = -JS/2-U$, while
for $J<0$ their position depends linearly on $J$ and also weakly
on $D$~\cite{Misiorny_PRB79/09}.
It is also interesting to note that in the P configuration the resonance peaks are almost exclusively
due to spin-up conductance $g^{\rm P}_\uparrow$, see Figs.~\ref{Fig:2}(c-f).
This is related to the fact that spin-up electrons
are the majority ones in both the left and right lead and thus tunneling of spin-up electrons
is favored, irrespective of $J$.
The explicit dependence of the conductance on $J$ is shown in Figs.~\ref{Fig:3}(a,b).

\emph{Tunnel magnetoresistance --} The effects associated with exchange
coupling $J$ are also pronounced in TMR,
$\textrm{TMR}=(g^\textrm{P}-g^\textrm{AP})/g^\textrm{AP}$ with
$g^{\rm P/AP}\equiv\sum_\sigma g^{\rm P/AP}_\sigma$, see
Figs.~\ref{Fig:2}(g)-(h). For $|J|\ll T_{\textrm{K}}$, the conductance in the
Kondo regime is generally larger in the AP configuration
than that in the P one, $g^{\rm AP}>g^{\rm P}$,
leading to negative TMR in major part of the Coulomb
blockade regime, except for $\varepsilon \approx -U/2$,
when $g^{\rm AP}<g^{\rm P}$, and TMR is positive.
When $|J|$ increases, the Kondo peak in
AP configuration becomes suppressed too, and positive
TMR in the blockade regime is restored. Since the suppression of
$g^{\rm AP}$ is more pronounced for $J<0$,
the corresponding TMR in the Kondo regime is larger for
$J<0$ than for $J>0$. Moreover, TMR for $J<0$ significantly
exceeds the corresponding Julliere's value, $2p^2/(1-p^2)$ ($=2/3$
for assumed parameters)~\cite{Julliere_PLA54/75}. On the other
hand, for empty or doubly occupied OL, where transport
is mainly determined by elastic cotunneling processes,
one always finds $g^{\rm P} > g^{\rm AP}$ with TMR approaching the Julliere's value.

The explicit variation of the spin-dependent conductance and the
corresponding TMR with the parameter $J$ is shown in
Fig.~\ref{Fig:3} for different values of $\varepsilon$.  When
$\varepsilon=-U/2$ and  $|J|\rightarrow0$, $\textrm{TMR}\approx
p^2/(1-p^2)$. In turn, as $|J|$ grows to energies corresponding to
the Kondo temperature, $|J|\approx T_{\rm K}$, TMR reaches a
minimum, where it takes negative values. Further growth of $|J|$
above $T_\textrm{K}$ is then accompanied by a significant increase
in TMR, especially for $J<0$. In the Coulomb blockade regime with
$\varepsilon\neq-U/2$, TMR is negative and constant for
$|J|<T_{\rm K}$, and starts increasing when $|J|>T_{\rm K}$ to
reach large positive values for $|J|\gg T_{\rm K}$ in the case of
$J<0$, see Fig.~\ref{Fig:3}(b). In turn, at resonance,
$\varepsilon=0$, TMR is positive and constant and only slightly
decreases when $|J| \approx \Gamma$, where a local minimum
develops due to the dependence of resonance energies on $J$.

\begin{figure}[t]
  \includegraphics[width=0.99\columnwidth]{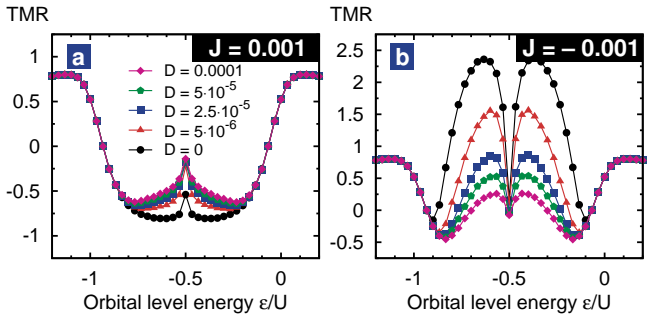}
  \caption{\label{Fig:4} (Color online)
  Influence of the molecule's magnetic anisotropy
  $D$ on TMR for the \emph{ferromagnetic} (a) and \emph{antiferromagnetic}
  (b) exchange coupling $J$.
  Parameters are the same as in Fig.~\ref{Fig:1} with $|J| = 2T_{\textrm{K}}$.}
\end{figure}

The uniaxial anisotropy $D$ modifies the energy spectrum and
electron states of MQD, and therefore  has a significant influence
on the Kondo state~\cite{Parks_Science328/10}, which can be
observed in the behavior of $g$ and TMR. In Fig.~\ref{Fig:4} we
show the $\varepsilon$-dependence of TMR for different values of the
anisotropy constant $D$. We note that the effects due to variation
of $D$ are mainly visible in the Kondo regime, while outside the
Coulomb blockade TMR is rather independent of $D$. Moreover, this
effect is more pronounced for $J<0$ than for $J>0$. This is
because magnetic anisotropy changes quantum states responsible for
the Kondo effect, suppressing the difference in TMR for $J<0$ and
$J>0$, see Fig.~\ref{Fig:4}(b).

\emph{Conclusions --} Using numerical renormalization group method
we have studied spin-dependent transport through a localized
orbital level coupled directly to ferromagnetic leads and
exchange-coupled to a magnetic core. We have shown that there is a
competition between interactions corresponding to two energy
scales: the Kondo temperature $T_{\textrm{K}}$ and exchange
coupling $J$. The linear conductance becomes suppressed when $|J|
\gtrsim T_{\textrm{K}}$ and this suppression is more pronounced in
the case of antiferromagnetic coupling $J$. Moreover, $J$ also has
a significant influence on TMR, which displays a nonmonotonic
dependence on $J$ with a minimum for $|J|\approx T_{\rm K}$  and
may be greatly enhanced when $J<0$ as compared to the case of
$J>0$.

%%%%%%%%%%%%%%%%%%%%%%%%%%%%%%%%%%%%%%%%%%%%%%%%%%%%%%%%%%%%%%%%
% acknowledgements
%%%%%%%%%%%%%%%%%%%%%%%%%%%%%%%%%%%%%%%%%%%%%%%%%%%%%%%%%%%%%%%%

M.M acknowledges the hospitality of Lehrstuhl J. von Delft. I.W.
acknowledges support from the Alexander von Humboldt Foundation
and funds of the Polish Ministry of Science and Higher Education
as research project for years 2008-2010.

%%%%%%%%%%%%%%%%%%%%%%%%%%%%%%%%%%%%%%%%%%%%%%%%%%%%%%%%%%%%%%%%
%%%%%%%%%%%%%%%%%%%%%%%%%%%%%%%%%%%%%%%%%%%%%%%%%%%%%%%%%%%%%%%%
%%%%%%%%%%%%%%%%%%%%%%%%%%%%%%%%%%%%%%%%%%%%%%%%%%%%%%%%%%%%%%%%

%\bibliographystyle{apsrev4-1}
%\bibliography{bibliography_PRL_NRG}

%merlin.mbs 2010-03-15 4.21a (PWD, AO, DPC)

\end{document}